\documentclass[prb,preprint,groupedaddress,showpacs,preprintnumbers,amsmath,amssymb]{revtex4}

\usepackage{graphicx}
\usepackage{dcolumn}
\usepackage{bm}

\begin{document}


\title{Electronic and optical properties of beryllium chalcogenides/silicon heterostructures}

\author{Titus Sandu }
 \email{titus.sandu@umontreal.ca}
\affiliation{D\'{e}partement de chimie\\
 Universit\'{e} de Montr\'{e}al \\ 
C.P. 6128, succursale Centre-ville, Montr\'{e}al, Qu\'{e}bec H3C 3J7, Canada
 }
\author{W. P. Kirk}%
 \email{kirk@uta.edu}
\affiliation{University of Texas at Arlington\\
  NanoFAB Center \\ Arlington, Texas 76019}%




\date{\today}

\begin{abstract}
We have calculated electronic and optical properties 
of Si/BeSe$_{0.41}$Te$_{0.59}$ heterostructures by a semiempirical 
$sp^{3}s^{*}$ tight-binding method. Tight-binding parameters and 
band bowing of BeSe$_{0.41}$Te$_{0.59}$ are considered through a recent 
model for highly mismatched semiconductor alloys. The band 
bowing and the measurements of conduction band offset lead to a type II 
heterostucture for Si/BeSe$_{0.41}$Te$_{0.59}$ with conduction band minimum 
in the Si layer and valence band maximum in the BeSe$_{0.41}$Te$_{0.59}$ layer.
The electronic structure and optical properties of various (Si$_{2})_{n 
}$/(BeSe$_{0.41}$Te$_{0.59})_{m}$ [001] superlattices have been considered. 
Two bands of interface states were found within the bandgap of bulk Si. Our 
calculations 
indicate that the optical edges are below the fundamental bandgap of 
bulk Si and the transitions are optically allowed.\end{abstract}

\maketitle


\section{Introduction}

\begin{figure}
\includegraphics{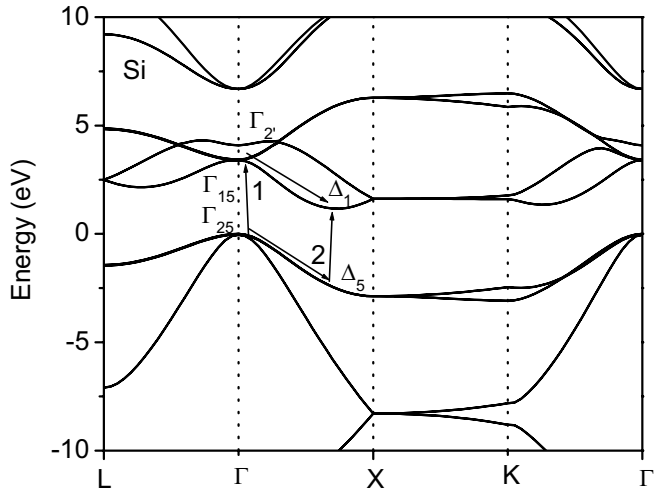}
\caption{\label{fig:1} The band structure of bulk Si with the parameters of 
\citeauthor{6} \cite{6} adapted with the spin-orbit constant \textit{$\Delta $} = 0.045 eV. The single 
group notations are used since the spin-orbit splitting cannot be distinguished 
on the chosen scale.}
\end{figure}

Silicon completely dominates the present
semiconductor market with the greatest shares of all sales. It has an ideal
bandgap (1.12 eV) for room temperature operation and its oxide 
(SiO$_{2})$ provides the necessary flexibility to fabricate millions of 
devices on a single chip. High integration implies high-speed operation that 
is limited by the interconnect propagation delay of the signal between 
devices. This constraint suggests that the integration of Si 
micro-electronics might be aided by optical interconnection. 
Unfortunately, 
silicon does not respond strongly to optical excitations because it is an 
indirect bandgap semiconductor: the band extrema for electrons and holes 
are located at different points in the Brillouin zone (Fig.~\ref{fig:1}). Therefore, 
intrinsic formation or recombination of electron-hole pairs becomes a 
three-particle event, which is weaker than a two-body process. Therefore, 
there are a number of areas where Si cannot compete
and this has allowed other materials to dominate such as optoelectronics 
where the indirect bandgap precludes
the use of Si to produce efficient light emitting diodes and lasers. 
If the performance of Si transistors
or circuits could be improved by addition of
another semiconductor material then numerous new
applications could open up. Silicon-germanium is one
such material which may be epitaxially grown on
silicon wafers and allows one to engineer the bandgap,
energy band structure, effective masses, mobilities
and numerous other properties while fabricating
circuits using conventional Si processing and tools.\cite{Paul99,Lee05}

There has been a tremendous effort in exploring the ways of breaking the silicon 
lattice symmetry and mixing different momentum states in order to induce 
optical gain. \cite{1,2} The radiative efficiency depends on the competition 
between non-radiative fast processes and relatively slow radiative 
processes. To optimize the efficiency we have to eliminate non-radiative 
channels by using high purity materials and increase the oscillator 
strength of radiative channels. The above criteria also apply to 
photovoltaics. One modality for improving the optical response is quantum confinement. 
Confinement of the charge density in quantum wells (QWs) permits the 
relaxation of the optical selection rules for an interband transition. In 
addition, the band folding in superlattice (SL) structures will enhance the 
absorption. Consequently a Si based SL will reduce the symmetry which 
translates into band folding toward the zone center; and as a result 
vertical transitions will be available at energies closer to the indirect 
bandgap. The discovery of visible light emission
from porous silicon has enticed many researchers
to reactivate studies of the optical properties
of silicon based nanostructures. \cite{Cullis97} Efficient visible light 
emission has been observed in Si-CaF$_{2}$ SLs,\cite{D'Avitaya95} 
Si-SiO$_{2}$ SLs,\cite{Lockwood96} Si nanopillar 
structures,\cite{Nassiopoulos95} and in Si crystallites.\cite{Kanemitsu93}

Beryllium chalcogenides have attracted much attention for the blue 
wavelength of region light emitting 
devices or UV detectors.\cite{Niiyama05,Kishino04} They provide long life-time 
II-VI lasers because of the strong covalent bonding of Be. 
BeTe and BeSe are wide-bandgap zinc blende semiconductors with lattice constants 
on either side of Si: they have the lattice constants of 
5.6269 and 5.1477 {\AA}, respectively, 3.6 {\%} larger and 5.2 {\%} smaller 
than Si. Vegard's law indicates that the lattice matched composition with Si 
is BeSe$_{0.41}$Te$_{0.59}$. Thus recent developments \cite{3,4} in the growth of 
silicon lattice-matched BeSe$_{0.41}$Te$_{0.59}$ open the opportunity for a 
new class of Si based devices.

In this paper we analyze quantum confinement and band folding in silicon based 
heterostructures (Si/BeSeTe heterostructures) as ways of enhancing light 
absorption. A semiempirical tight-binding (TB) method \cite{5} is used throughout the study. 
Although nowadays ab-initio methods (as they are formulated within density-functional theory)\cite{Jones89} 
are becoming routinely utilized for electronic structure and optical properties, 
they underestimate the bandgaps and therefore they need to be used with care. Moreover, the 
bandgap problem is solved within the so called GW (\emph{G = Green function}, \emph{W = screened Coulomb interaction}) 
approximation which is a computational 
demanding task and rarely used for large systems like ours (in our case there is an alloy plus a 
heterostructure).\cite{Aulbur00} In this work we choose a TB 
method together with a new method\cite{Sandu05} intended to deal with highly mismatched semiconductors, 
thus considering band-bowing of the BeSeTe alloy.
This paper is organized as follows. In section II we discuss methods to improve the light absorption in Si 
based heterostructures such as quantum confinement in a 
single (uncoupled) QW and band folding in SL structures. A significant 
increase in direct absorption is obtained for a narrow uncoupled silicon QW. 
In this spirit, we calculate the electronic and optical 
properties of Si/BeSeTe SLs using the TB method mentioned above. In the last 
section the conclusions are outlined. 

\section{Improved Optical Absorption by Quantum Confinement and Band Folding}

\subsection{Band Structure of BeSe$_{0.41}$Te$_{0.59}$}
Since both Si and BeSe$_{0.41}$Te$_{0.59}$ (as we will see below) are indirect band materials, full 
band calculations are required for a theoretical understanding of electron 
transport and optical properties in such heterostructures. For this purpose 
we use the empirical tight-binding (ETB) method as one of the most used 
tools in research of complex molecular and solid-state systems. \cite{8} 
Despite the fact that ETB is based on physical approximations such as the 
one-particle picture, short-range interactions, etc. (a posteriori justified, 
however), it gives fast and satisfactory results.  The short-range nature of the model is suitable for 
modeling heterointerfaces which are present in such quantum structures. 
One of the most popular TB models is the \textit{sp}$^{3}s^{*}$ model. \cite{6} This is a 
twenty-band model if spin-orbit coupling is included. \cite{7} The presence of an additional $s^{*}$ 
orbital is able to reproduce the bandgap in indirect 
semiconductors like silicon. The TB parameters of silicon are those from \citeauthor{6} 
\cite{6} augmented with spin-orbit coupling according to \citeauthor{7}. \cite{7} 

\begin{figure}
\includegraphics{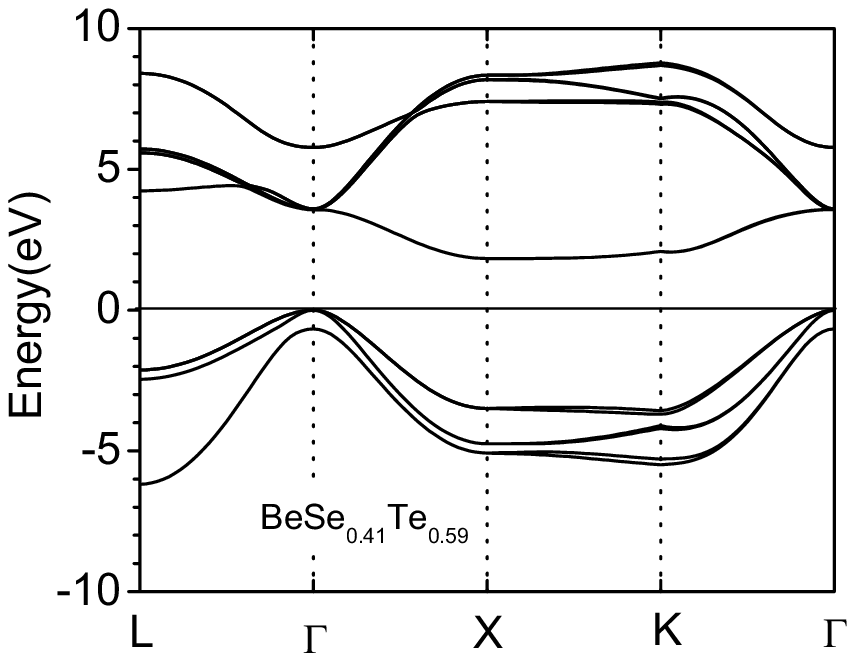}
\caption{\label{fig:2} Energy bands for BeSe$_{0.41}$Te$_{0.59}$ calculated with TB 
parameters from Ref.~\onlinecite{Sandu05}.}
\end{figure}

BeTe and BeSe are quite new materials in the sense that there are few 
experimental facts about these semiconductors. The bulk TB parameters of 
BeTe and BeSe were determined by fitting the GW  
calculations of Fleszar and Hanke. \cite{9} 
The band bowing has been included through a generalization of the band anti-crossing model for highly mismatched 
semiconductors. \cite{Sandu05} The details of the model and the TB parameters can be found in Ref.~\onlinecite{Sandu05}. 
The energy band diagram for BeSe$_{0.41}$Te$_{0.59}$ is 
shown in Fig.~\ref{fig:2}. The material is seen to be indirect bandgap with the 
conduction band minimum at the $X$ point. The predicted fundamental bandgap is 1.83 eV. The direct 
bandgap is 3.59 eV and the spin splitting is 0.67 eV.
Previous electrical measurements \cite{3,4} indicate a conduction band offset of at most 1.2 eV for 
the Si/BeSe$_{0.41}$Te$_{0.59}$ heterostructure. As we will see in the following subsections there are interface 
subbands induced at the heterostructure interfaces. This fact may overestimate the conduction band offset. 
Therefore we choose a value of 1 eV for the conduction band offset.
This makes the Si/BeSe$_{0.41}$Te$_{0.59}$ a type II heterostructure. 
Thus the top of the valence band in Si is higher than in the top of the valence band in  BeSe$_{0.41}$Te$_{0.59}$. 
The TB parameters and the conduction band offset will be used in the following 
sections in order to calculate electronic and optical properties of the 
Si/BeSe$_{0.41}$Te$_{0.59}$ SL. 

\subsection{A Single Quantum Well}

The absorption edges in bulk silicon are indirect and the jumping of the 
electron from valence band (VB) to conduction band (CB) is assisted by a 
phonon (in electron transitions from VB to CB the \textbf{\textit{k}} 
momentum must be conserved). The band structure of Si is shown schematically 
in the Fig.~\ref{fig:1}. The indirect bandgap is $\Delta _{1}-\Gamma 
_{25\mbox{'}}$. The absorption process involves two steps, one is 
electron-photon interaction and the other is electron-phonon interaction. 
Basically we have two processes denoted by 1 and 2. In process 1 an electron 
is first excited to the $\Gamma _{15}$ state and then by an emission of a 
phonon the electron arrives in the $\Delta _{1}$ CB state. Similarly for 
process 2, however, in this process it starts with an emission of a phonon. 
For this reason, there is a long absorption tail 
between 1.12 eV and and about 3 eV that reflects the indirect nature of the 
bandgap. Then a sharp rise in absorption occurs starting around 3.2 eV (380 nm) 
that is associated with the direct transition at 
$\Gamma $ point ($\Gamma _{25\mbox{'}} \quad  \to  \quad \Gamma _{15})$ whose 
energy is 3.4 eV (365 nm). 

In direct bandgap materials light absorption in 2D systems is formally 
similar to 3D systems unlike the light absorption in indirect bandgap systems. 
The absorption coefficient for an uncoupled Si/ BeSe$_{0.41}$Te$_{0.59}$ QW has 
basically three components \cite{11}

\begin{widetext}
\begin{eqnarray}
\label{eq:2}
\alpha \left( \omega \right) = A\;\left[ \sum\limits_{eh,\lambda_{q}} {p_{eh}^d\delta 
\left( {E_g + E_e + E_h - \hbar \omega } \right)} + \sum\limits_{eh,\lambda_
{q}} {p_{eh}^a\,n_{\lambda_{q}} \,\delta \left( {E_g + E_e + E_h - \hbar 
\omega - \hbar \omega _{\lambda_{q}} } \right)}\right.\nonumber\\ 
 \left. { + \sum\limits_{eh,\lambda_{q}} {p_{eh} ^e\left( {n_{\lambda q} + 
1} \right)\,\,\delta \left( {E_g + E_e + E_h - \hbar \omega + \hbar \omega 
_{\lambda q} } \right)} } \right],
\end{eqnarray}
\end{widetext}

\noindent
where $A$ is a constant, $p_{eh}^{d}$ gives the direct bandgap contribution 
and $p_{eh}^{a}$ gives the phonon assisted contribution with 1-phonon 
absorption, and $p_{eh}^{e}$ is the phonon assisted contribution with 
1-phonon emission. Basically,

\begin{equation}
\label{eq:3}
p_{eh} ^d = p_{cv} I_{eh} \left( {\bm{\mbox{k}}_0 } \right),
\end{equation}

\begin{equation}
\label{eq:4}
p_{eh,\lambda \bm{q}} ^e = p_{eh,\lambda \bm{q}} ^a = p_{cv} I_{eh} 
\left( {\bm{\mbox{q}}\; - \;\bm{\mbox{k}}_0 } \right)R_\lambda. 
\end{equation}

\begin{figure}
\includegraphics{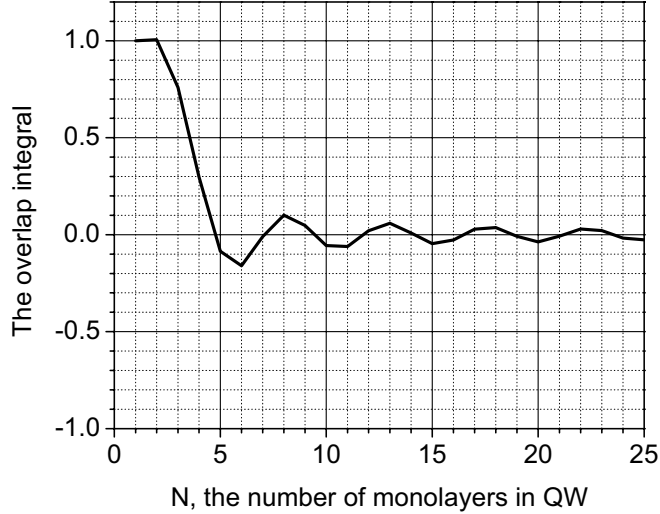}
\caption{\label{fig:3} The overlap integral (Eq.(\ref{eq:5}) for infinite wells in silicon. The 
wave vector \textbf{k}$_{0}$ in Eq.(\ref{eq:5}) is the wave vector for 
silicon conduction valleys along $z-$axis.}
\end{figure}

\noindent
$p_{cv}$ is the bulk dipole matrix element between bands, $R_{\lambda }$ is 
the matrix element contribution from electron-phonon interaction, and 
$I_{eh}$ is the overlap between the envelope-functions at different Brillouin 
points:

\begin{equation}
\label{eq:5}
I_{eh} \left( \bm{\mbox{q}} \right) = \int {d\;\bm{\mbox{r}}\;\psi _e ^{\ast} \left( 
\bm{\mbox{r}} \right)\,} \psi _h \left( \bm{\mbox{r}} \right)\;e^{ - 
i \bm{\mbox{k}}_{0}\bm{\mbox{r}}},
\end{equation}

\noindent
where $\bm{\mbox{k}}$$_{0}$ is the location in $k-$space of each conduction 
valley. For silicon, if $z$ is the growth direction, then the 2 valleys along 
z-directions (with $k_{y} = k_{x} = 0 )$ are responsible for direct transitions. The 
other four valleys contribute to phonon-assisted absorption. For infinite 
wells the overlap integral $I_{eh}$ has the behavior depicted in Fig.~\ref{fig:3}. For 
very narrow QW's the overlap integral tends to one, which means that in a 
genuine 2D system only the transverse momentum has to be conserved. For the 
other asymptotic limit, i.e. very wide well, the overlap integral vanishes. 
In physical terms, this says that the system became genuinely 3D and any 
electromagnetic transition has to be vertical in the absence of phonons. 
Moreover, from Fig.~\ref{fig:3} we may expect to have a strong direct transition (with 
an overlap integral no less than 0.1) for a QW with a width up to 20 {\AA}, 
i.e. an ultrathin QW. In the following we will discuss the possibility of 
improving the oscillator strength in Si/ BeSe$_{0.41}$Te$_{0.59}$ based SLs. 

\subsection{Electronic Structures of 
\textbf{(Si}$_{2}$\textbf{)}$_{n}$\textbf{/(BeSe}$_{0.41}$\textbf{Te}$_{0.59}$\textbf{)}$_{m}$\textbf{ [001]}
Superlattices}

The use of SL structures relaxes the condition of ultrathin QW. The alternation of QWs and 
barriers along [001] growth 
axis will generate a band folding in the SL and band mixing of zone-center 
and zone edge states. \cite{12,13} The band mixing will inherently enhance the 
oscillator strength for direct transition in Si structures. On the other 
hand, band folding will induce states in which vertical transitions are 
possible at energies far lower than 3.4 eV, the lowest energy for 
vertical transitions in bulk Si. 

\begin{figure*}
\includegraphics{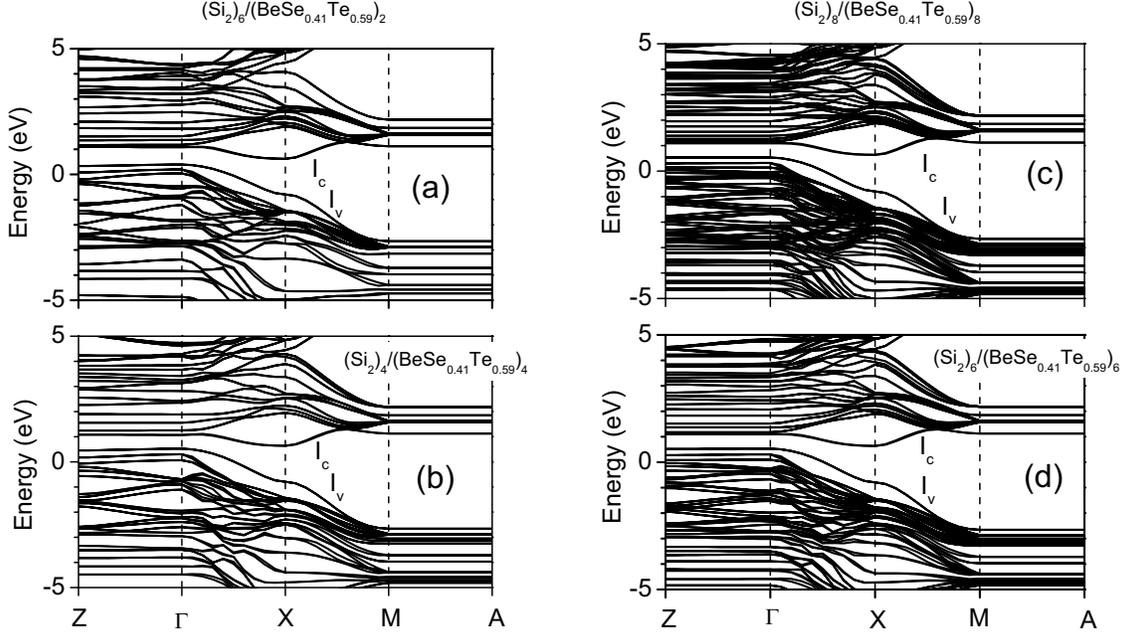}
\caption{\label{fig:4} Band structures of (Si$_{2})_{n}$/(BeSe$_{0.41}$Te$_{0.59})_{m}$ 
[001] superlattices for (a) $n = 6, m=2$, (b) $n = m = 4$, (c) $n = m = 8$, and (d) $n = m = 6$. 
The interface bands are I$_{v}$ and I$_{c}$.}
\end{figure*}

We consider a Si/BeSe$_{0.41}$Te$_{0.59}$ SL whose layers are perpendicular 
to [001] direction. We employ the nearest neighbor \textit{sp}$^{3}s^{*}$ TB Hamiltonian 
including spin-orbit interaction. Although there are optimized nearest-neighbor TB 
parameters in the 
literature\cite{Klimeck00} we chose the parameters given in Ref.~\onlinecite{6}. These parameters 
include the chemical trends which are needed for an overall description 
of both valence and conduction bands. As it was noted before, the optimized 
nearest-neighbor parameters may have unphysical interaction energies and they 
are dedicated for either valence band or conduction band calculations.\cite{Klimeck00} 
Moreover the nearest-neighbor model has the inability to reproduce the 
transverse/light X-valley effective mass. However, the conduction band 
edges in low-dimensional structures will be determined by the 
longitudinal/heavy effective mass at the conduction band minimum. Therefore 
we opted for a less accurate yet a very representative set of parameters 
because these parameters are within the physical capabilities of the nearest 
neighbor model.

We denote this SL as a (Si$_{2})_{n 
}$/(BeSe$_{0.41}$Te$_{0.59})_{m}$ SL with $n$ two-atom thick layers of Si and 
$m$ two-atom thick layers of BeSe$_{0.41}$Te$_{0.59}$ repeated periodically. 
The defined supercell consists of 2 ($n+m)$ adjacently bonded atoms Si, Si, 
\ldots ..Si, Be, Te/Se, Be, Te/Se, \ldots ..Be, Te/Se. 
Covalent bonds between Si on the one hand and Be or Se/Te on the other  are 
not electrically neutral.\cite{Harrison78}  For Si-based heterostructures special care has 
to be exercised due to excess/deficiency of electrons at interface and due 
to antiphase domains.\cite{Kroemer87}  These problems can be overcome by vicinal growth and 
passivated surfaces.\cite{3,4}  In this way one can minimize the interface charging 
effects.  Moreover, self-consistent ab-initio calculations show that 
the electric fields due to difference in ionicity between the two 
semiconductor constituents of a superlattice dominate the total electric 
field.\cite{Eppenga89} Therefore the accumulated charge with respect to the bulk  
contributes in a much lesser degree to the total electric field in 
the superlattice structure.  In addition to that, the ETB fully 
self-consistent calculations as those described by Della Sala and coworkers\cite{Sala99}
 are not much different from those non-self-consistent calculations 
provided that the current densities in the confined system are low.\cite{Bonfiglio00}  All the 
above facts lead us to conclude that a "zero-field" approximation will 
give us at least a qualitative picture of Si/BST structures.   

\begin{figure*}
\includegraphics{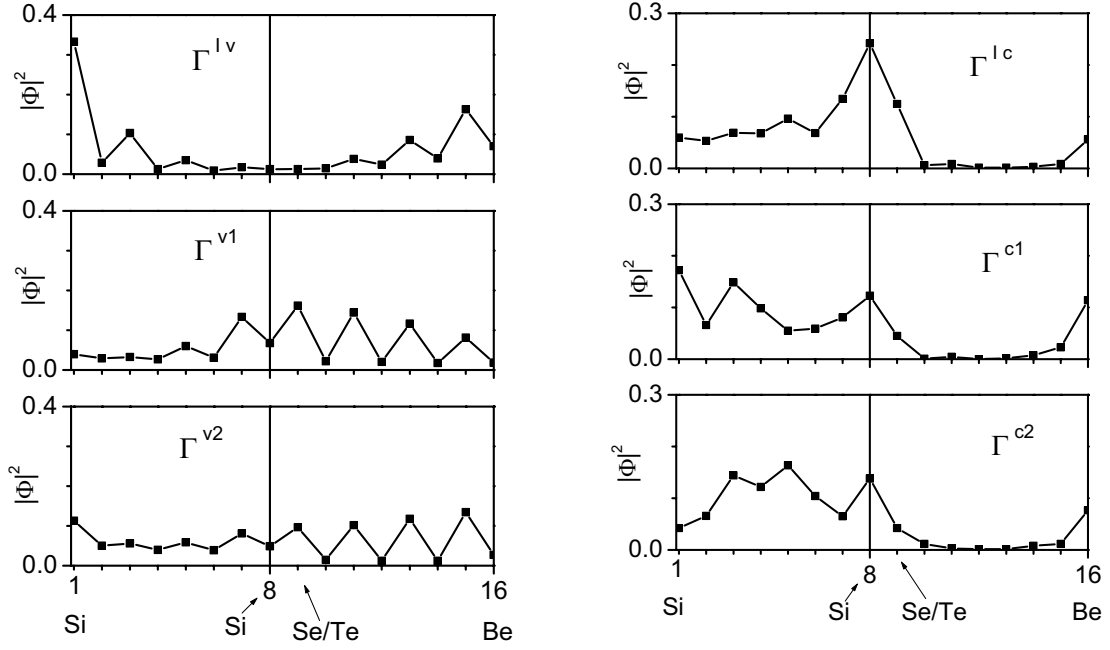}
\caption{\label{fig:5}Planar charge densities in the 
(Si$_{2})_{4}$/(BeSe$_{0.41}$Te$_{0.59})_{4}$ [001] superlattice at 
\textit{$\Gamma $} point. The occupied states are the interface state $\Gamma ^{\,I_V }$ and the 
first top two confined states ($\Gamma ^{\,V1}$ and $\Gamma ^{\,V2})$. 
The empty states are the interface state $\Gamma ^{\,I_c }$ and the first two 
confined states ($\Gamma ^{\,C1}$ and $\Gamma ^{\,C2})$. The solid vertical 
lines denote the interfaces.}
\end{figure*}

The TB matrix 
elements of the Si/BeSe$_{0.41}$Te$_{0.59}$ SL are taken directly from 
those bulk values. The on-site energies of the BeSe$_{0.41}$Te$_{0.59}$ are 
accordingly changed to match the valence band offset at the interface. 
Simple averages were used to supply the parameters connecting different 
materials at the interface. Since the spin orbit must be included, the SL 
Hamiltonian will be represented as having $20\,\left( {n + m} \right)$ 
functions. Once the TB matrix elements have been established the SL band 
structure reduces to the diagonalization of the $20\,\left( {n + m} 
\right)\times 20\,\left( {n + m} \right)$ matrix Hamiltonian. The band 
structure of (Si$_{2})_{n }$/(BeSe$_{0.41}$Te$_{0.59})_{m}$ SL for $m = 2$ 
and $n = 6$, $m = n = 4$, $m = n = 8$, and $m = n = 6$ are displayed in Fig.~\ref{fig:4}. 
The zero of energy corresponds to the top of the valence band in bulk Si. 
The band folding effect can be seen as many crowded subbands. Two interface 
bands (I$_{v}$ and I$_{c})$, one empty and one occupied, were found. They lie 
in the lower and upper parts of the bandgap of bulk silicon, respectively. 
The origin of these interface bands rests on the polar nature of the 
interface as was also found in GaAs/Ge SL. \cite{14,Laref03,Laref06} The polarity of the 
interface originates from the large differences in the on-site energies for 
the constituent atoms (Si and Be or Se/Te). Even if a (110) non-polar 
interface is used, one interface band is still found in II-VI/IV SLs. \cite{15,16} 
We calculated the planar charge density of some of the occupied and empty 
band edge states of the SL with $\mbox{m} = \mbox{n} = \mbox{4}$ at the $\Gamma 
$ point in the Brillouin zone. The planar charge density of some of the band 
edge states and interface states are depicted in Fig.~\ref{fig:5}. 
By interface states we mean states that 
die away within a few layers from the interface. We denote by $\Gamma ^{I\,v}$, 
the interface states of the $I_{v}$ band and by $\Gamma ^{I\,c}$, the 
interface states of the $I_{c}$ band at $\Gamma $ point. The charge density 
of the $\Gamma ^{I\,v}$ interface state has a maximum at the Be-Si 
interface, while the next occupied state is more confined in the 
BeSe$_{0.41}$Te$_{0.59}$ layer, but it still spills over into the Si barrier.  
The third occupied state spills more over into the Si slab, not only due to interface 
but also due to the fact that the state is closer to the top of hole barrier in Si. 
The charge density of the $\Gamma ^{I\,c}$ 
interface state has a maximum at the Si-Se/Te interface, while the next 
empty state is confined, but it is still localized toward the interfaces. 
Only the third empty state is genuinely confined to the silicon slab. We 
believe that the charge distribution of the interface subbands is related to 
the acceptor behavior of Be and donor 
behavior of Se and Te with respect to Si. \cite{17} 

\subsection{ Optical Properties of 
\textbf{(Si}$_{2}$\textbf{)}$_{n}$\textbf{/(BeSe}$_{0.41}$\textbf{Te}$_{0.59}$\textbf{)}$_{m}$\textbf{ [001]}
Superlattices}

The electronic contribution to absorption spectrum is given by $\sigma 
_{abs} \left( \omega \right)\sim \omega \,\varepsilon _2 \left( \omega 
\right)$, where $\sigma _{abs} \left( \omega \right)$ is the absorption 
coefficient and \cite{8}

\begin{equation}
\label{eq:6}
\varepsilon _2 \left( \omega \right) = \frac{2\pi ^2\hbar \,e^2}{m\omega 
\,\Omega }\sum\limits_{c,v,k} {f_{cv,k} \delta \left( {E_{c,k} - E_{v,k} - 
\hbar \omega } \right)} 
\end{equation}

\noindent
is the imaginary part of the dielectric function. Here $m$ is the electron mass, 
\textit{$\Omega $} is the volume, $e$ is the electron charge, $\hbar $ is the Planck constant and 
$f_{cv,k} $ is the oscillator strength for the direct transition from the 
state $\left| {v,{\rm {\textbf k}}} \right\rangle $ to $\left| {c,{\rm {\textbf k}}} 
\right\rangle $, with the photon momentum neglected. In Eq. (\ref{eq:6}) 
it is invoked the electric dipole approximation in the low-temperature 
case. In ETB the expressions for optical absorption and, in general, the 
interaction with the electromagnetic field are different than those 
encountered in conventional quantum mechanics due to incompleteness of the 
basis.\cite{Boykin95}  In the electric dipole approximation, however, the expressions of 
optical absorption are the same. The full temperature-dependent equations 
for optical absorption are derived in Refs.~\onlinecite{18} and \onlinecite{Boykin02}.  
Beyond the dipole 
approximation the standard expression (\ref{eq:6}) is no longer valid.  
The problem is treated exhaustively in Ref. \onlinecite{Boykin02b}, where it is 
explicitly shown how the Hamiltonian is altered by the vector 
potential of the electromagnetic field. 

The oscillator strength 
is defined as

\begin{equation}
\label{eq:7}
f_{cv,k} = \frac{2}{m}\frac{\left| {\left\langle {c,{\rm {\textbf k}}} 
\right|\,{\rm {\bm \varepsilon }} \cdot {\rm {\textbf p}}\,\left| {v,{\rm {\textbf 
k}}} \right\rangle } \right|^2}{E_{c,k} - E_{v,k} } .
\end{equation}

\noindent
In Eq. (\ref{eq:7}) $\left| {v,{\rm {\textbf k}}} \right\rangle $ and $\left| {c,{\rm 
{\textbf k}}} \right\rangle $ are the valence and conduction band eigenstates, 
$E_{v,k}$ and $E_{c,k}$ are their corresponding energies, ${\rm {\bm 
\varepsilon }}$ is the polarization of light, and \textbf{p} is the 
momentum operator. In the empirical tight-binding approach the momentum 
matrix element is defined as \cite{18}

\begin{equation}
\label{eq:8}
\left\langle {c,{\rm {\textbf k}}} \right|{\rm {\textbf p}}\left| {c,{\rm {\textbf k}}} 
\right\rangle = \frac{m}{\hbar }\left\langle {c,{\rm {\textbf k}}} 
\right|\,\nabla _{\rm {\textbf k}} H\left( {\rm {\textbf k}} \right)\,\left| {v,{\rm 
{\textbf k}}} \right\rangle .
\end{equation}

\begin{figure}
\includegraphics{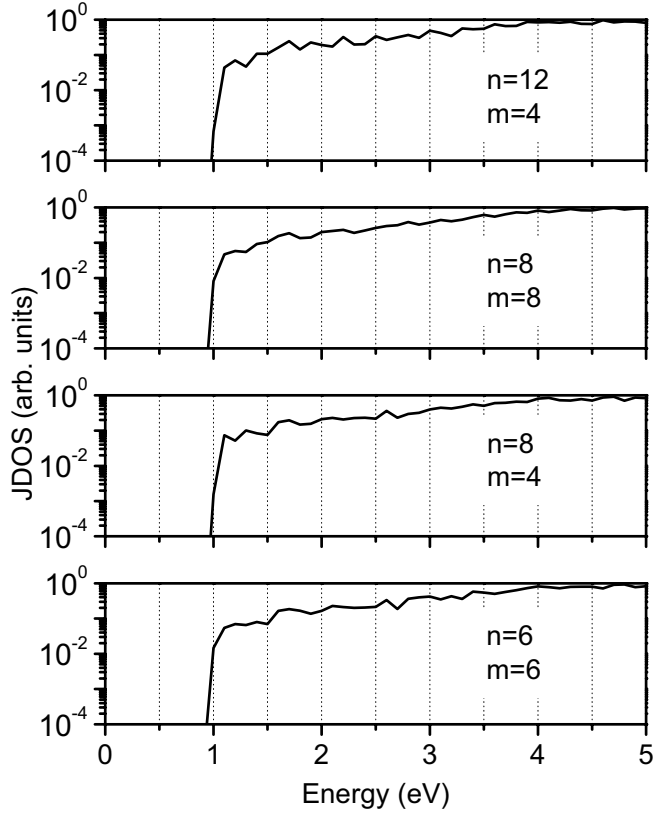}
\caption{\label{fig:6} Joint density of states corresponding to optical transitions of 
(Si$_{2})_{n}$/(BeSe$_{0.41}$Te$_{0.59})_{m}$ [001] superlattices with $n = 12$ and $m = 4$, 
$n = m = 8$, $n = 8$ and $m = 6$, and $m = n = 6$.}
\end{figure}

Although Eq.~(\ref{eq:8}) does not consider the intra-atomic part of the 
momentum matrix element,\cite{Pedersen01,21} it is gauge invariant.\cite{18,Boykin02} 
Its accuracy can be improved by going beyond nearest neighbor interaction or by using 
non-orthogonal orbitals.\cite{Sandu2005}
We first calculate the joint densities of states (JDOS) because the variation of the 
oscillator strength over the Brillouin zone is small. \cite{13} We 
assume that light is propagating along the SL growth direction. The JDOS 
represents the number of states that can undergo energy and 
\textbf{k}-conserving transitions for photon frequencies between 
\textit{$\omega $} and $\omega  + d\omega $. The JDOS associated with 
Eq. (\ref{eq:6}) are shown in Fig.~\ref{fig:6} for SLs with $n = 12$ and $m = 4$, $n = m = 8$, 
$n = 8$ and $m = 4$, 
and $m = n = 6$. A 0.05 eV broadening was considered for each electronic energy. 
The summation over Brillouin zone was replaced by the summation over special points 
in the Brillouin zone. \cite{19,20} 

\begin{figure}
\includegraphics{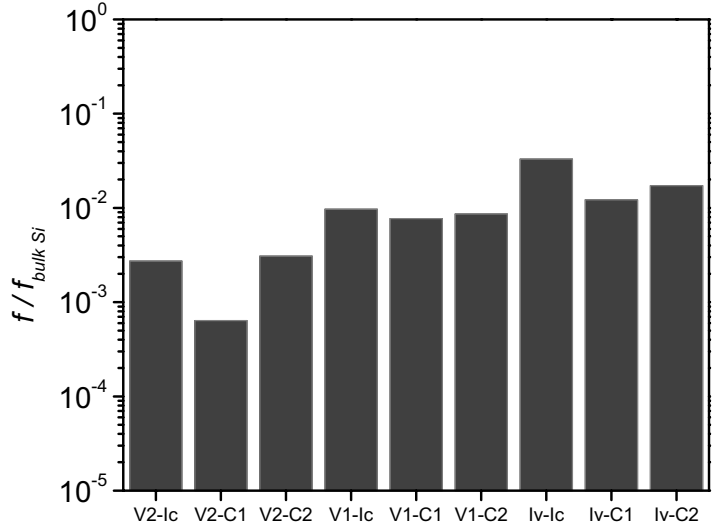}
\caption{\label{fig:7} The oscillator strengths $f$ of several interband transitions relative 
to the direct transition in bulk Si ($\Gamma _{25\mbox{'}}  \quad  \to $ 
$\Gamma _{15}) \quad f_{bulkSi}$ for the 
(Si$_{2})_{4}$/(BeSe$_{0.41}$Te$_{0.59})_{4}$ [001] superlattice.}
\end{figure}

Due to band folding the absorption edges for vertical 
transitions are lowered toward the indirect bandgap of bulk Si. Moreover, 
the absorption edges are not determined by the interface subbands but by the energy difference 
between the lowest empty confined state and highest occupied confined state. Thus the absorption edges for vertical 
transitions are below the indirect bandgap of Si because the heterostructure 
is of type II. The curves rise slowly and, with increasing $m\left( { n} \right)$, the 
absorption edges extend to lower energies. However, the thickness of the BeSe$_{0.41}$Te$_{0.59}$ 
layers play a greater role than the thickness of Si layers in the variation of the absorption edge. It 
can be explained by the difference of the electron effective masses in Si and hole effective masses in 
BeSe$_{0.41}$Te$_{0.59}$, i.e., the heavy hole effective mass in BeSe$_{0.41}$Te$_{0.59}$ is smaller than the 
effective mass of heavy electron valley in Si. Similar results were found in the 
calculations for porous Si with periodic boundary conditions:\cite{21} the absorption band edge does not vary 
much with Si porosity. 
The strength of the optical absorption is also 
determined by the oscillator strength, thus we can also check if the transition are allowed 
or not allowed. We denote by Iv, V1, and V2 the first 
three top valence subbands, and by Ic, C1, and C2 the first three conduction 
subbands. We calculated the oscillator strengths of several interband 
transitions relative to the oscillator strength of the direct transition in 
bulk Si. The results are shown in Fig.~\ref{fig:7} for $m = n = 4$. The oscillator 
strengths of interband transitions are at least 10 times smaller than their 
bulk counterpart and basically range from 10 $^{\mbox{--}3}$ to 10 $^{ - 1}$ 
relative to $\Gamma _{25\mbox{'}}\quad\to\quad\Gamma _{15}$ transition in 
bulk Si. The strongest transitions are those coming from the interface subband and those coming from the 
first confined 
hole level to the confined electron levels. 

\section{Conclusions}
We studied the electronic and optical properties of silicon-based heterostructures 
(Si/BeSe$_{0.41}$Te$_{0.59})$. 
Based on the fact that significant light absorption occurs for thin uncoupled  Si quantum wells 
we studied the 
(Si)$_{n}$/(BeSe$_{0.41}$Te$_{0.59})_{ m}$ superlattice structures with a 
semiempirical tight-binding method 
in the \textit{sp}$^{3}s^{*}$ nearest neighbor model including spin-orbit interaction. 
The tight-binding model considered large band bowing in the BeSe$_{0.41}$Te$_{0.59}$ alloy. 
The heterostructure Si/BeSe$_{0.41}$Te$_{0.59}$ was found to be of type II.
Electronic structure and optical properties were calculated for various 
superlattice structures. Two interface bands were found to exist in the bandgap 
of bulk silicon. We found that band folding induces vertical transitions 
below the indirect bandgap of bulk Si. In addition, calculated oscillator 
strengths for vertical transitions near the optical band edge show the 
mixing of the zone-center and zone edge states of bulk Si for conduction 
subbands. Therefore, the transitions are optically allowed and the response 
of silicon based heterostructures to illumination is enhanced.

\begin{acknowledgments}
This work was supported in part by NASA grant NCC-1-02038  and by the Texas 
Advanced Technology program under grant No. 003656-0029-2003.
The work has been also supported in part by NSERC grants no. 311791-05 and 315160-05.
One of the authors (T.S.) wishes to acknowledge generous support in the form of computer resources from
the R\'{e}seau Qu\'{e}b\'{e}cois de Calcul de Haute Performance.
\end{acknowledgments}




\begin{thebibliography}{45}
\expandafter\ifx\csname natexlab\endcsname\relax\def\natexlab#1{#1}\fi
\expandafter\ifx\csname bibnamefont\endcsname\relax
  \def\bibnamefont#1{#1}\fi
\expandafter\ifx\csname bibfnamefont\endcsname\relax
  \def\bibfnamefont#1{#1}\fi
\expandafter\ifx\csname citenamefont\endcsname\relax
  \def\citenamefont#1{#1}\fi
\expandafter\ifx\csname url\endcsname\relax
  \def\url#1{\texttt{#1}}\fi
\expandafter\ifx\csname urlprefix\endcsname\relax\def\urlprefix{URL }\fi
\providecommand{\bibinfo}[2]{#2}
\providecommand{\eprint}[2][]{\url{#2}}

\bibitem[{\citenamefont{Paul}(1999)}]{Paul99}
\bibinfo{author}{\bibfnamefont{D.~J.} \bibnamefont{Paul}},
  \bibinfo{journal}{Adv. Materials} \textbf{\bibinfo{volume}{11}},
  \bibinfo{pages}{191} (\bibinfo{year}{1999}).

\bibitem[{\citenamefont{Lee and Fitzgerald}(2005)}]{Lee05}
\bibinfo{author}{\bibfnamefont{M.~L.} \bibnamefont{Lee}} \bibnamefont{and}
  \bibinfo{author}{\bibfnamefont{E.~A.} \bibnamefont{Fitzgerald}},
  \bibinfo{journal}{J. Appl. Phys.} \textbf{\bibinfo{volume}{97}},
  \bibinfo{pages}{011101} (\bibinfo{year}{2005}).

\bibitem[{\citenamefont{Iyer and Xie}(1993)}]{1}
\bibinfo{author}{\bibfnamefont{S.~S.} \bibnamefont{Iyer}} \bibnamefont{and}
  \bibinfo{author}{\bibfnamefont{Y.~H.} \bibnamefont{Xie}},
  \bibinfo{journal}{Science} \textbf{\bibinfo{volume}{260}},
  \bibinfo{pages}{40} (\bibinfo{year}{1993}).

\bibitem[{\citenamefont{Tsu}(1993)}]{2}
\bibinfo{author}{\bibfnamefont{R.}~\bibnamefont{Tsu}},
  \bibinfo{journal}{Nature} \textbf{\bibinfo{volume}{364}},
  \bibinfo{pages}{338} (\bibinfo{year}{1993}).

\bibitem[{\citenamefont{Cullis et~al.}(1997)\citenamefont{Cullis, Canham, and
  Calcott}}]{Cullis97}
\bibinfo{author}{\bibfnamefont{A.~G.} \bibnamefont{Cullis}},
  \bibinfo{author}{\bibfnamefont{L.~T.} \bibnamefont{Canham}},
  \bibnamefont{and} \bibinfo{author}{\bibfnamefont{P.~D.~J.}
  \bibnamefont{Calcott}}, \bibinfo{journal}{J. Appl. Phys.}
  \textbf{\bibinfo{volume}{82}}, \bibinfo{pages}{909} (\bibinfo{year}{1997}).

\bibitem[{\citenamefont{DAvitaya et~al.}(1995)\citenamefont{DAvitaya,
  Vervoort, Ossicini, A.Fasolino, and Bernardini}}]{D'Avitaya95}
\bibinfo{author}{\bibfnamefont{F.~A.} \bibnamefont{DAvitaya}},
  \bibinfo{author}{\bibfnamefont{L.}~\bibnamefont{Vervoort}},
  \bibinfo{author}{\bibfnamefont{S.}~\bibnamefont{Ossicini}},
  \bibinfo{author}{\bibnamefont{A.Fasolino}}, \bibnamefont{and}
  \bibinfo{author}{\bibfnamefont{F.}~\bibnamefont{Bernardini}},
  \bibinfo{journal}{Europhys. Lett.} \textbf{\bibinfo{volume}{31}},
  \bibinfo{pages}{25} (\bibinfo{year}{1995}).

\bibitem[{\citenamefont{Lockwood et~al.}(1996)\citenamefont{Lockwood, Lu, and
  Baribeau}}]{Lockwood96}
\bibinfo{author}{\bibfnamefont{D.}~\bibnamefont{Lockwood}},
  \bibinfo{author}{\bibfnamefont{Z.}~\bibnamefont{Lu}}, \bibnamefont{and}
  \bibinfo{author}{\bibfnamefont{J.-M.} \bibnamefont{Baribeau}},
  \bibinfo{journal}{Phys. Rev. Lett.} \textbf{\bibinfo{volume}{76}},
  \bibinfo{pages}{539} (\bibinfo{year}{1996}).

\bibitem[{\citenamefont{Nassiopoulos et~al.}(1996)\citenamefont{Nassiopoulos,
  Grigoropoulos, Papadimitiriu, and Gogolides}}]{Nassiopoulos95}
\bibinfo{author}{\bibfnamefont{A.}~\bibnamefont{Nassiopoulos}},
  \bibinfo{author}{\bibfnamefont{S.}~\bibnamefont{Grigoropoulos}},
  \bibinfo{author}{\bibfnamefont{D.}~\bibnamefont{Papadimitiriu}},
  \bibnamefont{and}
  \bibinfo{author}{\bibfnamefont{E.}~\bibnamefont{Gogolides}},
  \bibinfo{journal}{Phys. Stat. Sol. (b)} \textbf{\bibinfo{volume}{190}},
  \bibinfo{pages}{91} (\bibinfo{year}{1996}).

\bibitem[{\citenamefont{Kanemitsu et~al.}(1993)\citenamefont{Kanemitsu, Uto,
  Masumoto, Matsumoto, Futagi, and Mimura}}]{Kanemitsu93}
\bibinfo{author}{\bibfnamefont{Y.}~\bibnamefont{Kanemitsu}},
  \bibinfo{author}{\bibfnamefont{H.}~\bibnamefont{Uto}},
  \bibinfo{author}{\bibfnamefont{Y.}~\bibnamefont{Masumoto}},
  \bibinfo{author}{\bibfnamefont{T.}~\bibnamefont{Matsumoto}},
  \bibinfo{author}{\bibfnamefont{T.}~\bibnamefont{Futagi}}, \bibnamefont{and}
  \bibinfo{author}{\bibfnamefont{H.}~\bibnamefont{Mimura}},
  \bibinfo{journal}{Phys. Rev. B} \textbf{\bibinfo{volume}{48}},
  \bibinfo{pages}{2827} (\bibinfo{year}{1993}).

\bibitem[{\citenamefont{Niiyama and Watanabe}(2005)}]{Niiyama05}
\bibinfo{author}{\bibfnamefont{Y.}~\bibnamefont{Niiyama}} \bibnamefont{and}
  \bibinfo{author}{\bibfnamefont{M.}~\bibnamefont{Watanabe}},
  \bibinfo{journal}{Semicond. Sci. Technol.} \textbf{\bibinfo{volume}{20}},
  \bibinfo{pages}{1187} (\bibinfo{year}{2005}).

\bibitem[{\citenamefont{Kishino and Nomura}(2004)}]{Kishino04}
\bibinfo{author}{\bibfnamefont{K.}~\bibnamefont{Kishino}} \bibnamefont{and}
  \bibinfo{author}{\bibfnamefont{I.}~\bibnamefont{Nomura}},
  \bibinfo{journal}{Phys. Stat. Sol. (c)} \textbf{\bibinfo{volume}{1}},
  \bibinfo{pages}{1477} (\bibinfo{year}{2004}).

\bibitem[{\citenamefont{Clark et~al.}(2000)\citenamefont{Clark, Maldonado,
  Barrios, Spencer, Bate, and Kirk}}]{3}
\bibinfo{author}{\bibfnamefont{K.}~\bibnamefont{Clark}},
  \bibinfo{author}{\bibfnamefont{E.}~\bibnamefont{Maldonado}},
  \bibinfo{author}{\bibfnamefont{P.}~\bibnamefont{Barrios}},
  \bibinfo{author}{\bibfnamefont{G.~F.} \bibnamefont{Spencer}},
  \bibinfo{author}{\bibfnamefont{R.~T.} \bibnamefont{Bate}}, \bibnamefont{and}
  \bibinfo{author}{\bibfnamefont{W.~P.} \bibnamefont{Kirk}},
  \bibinfo{journal}{J. Appl. Phys.} \textbf{\bibinfo{volume}{88}},
  \bibinfo{pages}{7201} (\bibinfo{year}{2000}).

\bibitem[{\citenamefont{Kirk et~al.}(2000)\citenamefont{Kirk, Clark, Maldonado,
  Basit, Bate, and Spencer}}]{4}
\bibinfo{author}{\bibfnamefont{W.~P.} \bibnamefont{Kirk}},
  \bibinfo{author}{\bibfnamefont{K.}~\bibnamefont{Clark}},
  \bibinfo{author}{\bibfnamefont{E.}~\bibnamefont{Maldonado}},
  \bibinfo{author}{\bibfnamefont{N.}~\bibnamefont{Basit}},
  \bibinfo{author}{\bibfnamefont{R.~T.} \bibnamefont{Bate}}, \bibnamefont{and}
  \bibinfo{author}{\bibfnamefont{G.~F.} \bibnamefont{Spencer}},
  \bibinfo{journal}{Supperlatt. Microstruct.} \textbf{\bibinfo{volume}{28}},
  \bibinfo{pages}{377} (\bibinfo{year}{2000}).

\bibitem[{\citenamefont{Vogl et~al.}(1983)\citenamefont{Vogl, Hjalmarson, and
  Dow}}]{6}
\bibinfo{author}{\bibfnamefont{P.}~\bibnamefont{Vogl}},
  \bibinfo{author}{\bibfnamefont{H.~P.} \bibnamefont{Hjalmarson}},
  \bibnamefont{and} \bibinfo{author}{\bibfnamefont{J.~D.} \bibnamefont{Dow}},
  \bibinfo{journal}{J. Phys. Chem. Solids} \textbf{\bibinfo{volume}{44}},
  \bibinfo{pages}{365} (\bibinfo{year}{1983}).

\bibitem[{\citenamefont{Slater and Koster}(1954)}]{5}
\bibinfo{author}{\bibfnamefont{J.~C.} \bibnamefont{Slater}} \bibnamefont{and}
  \bibinfo{author}{\bibfnamefont{G.~F.} \bibnamefont{Koster}},
  \bibinfo{journal}{Phys. Rev.} \textbf{\bibinfo{volume}{94}},
  \bibinfo{pages}{1498} (\bibinfo{year}{1954}).

\bibitem[{\citenamefont{Jones and Gunnarsson}(1989)}]{Jones89}
\bibinfo{author}{\bibfnamefont{R.}~\bibnamefont{Jones}} \bibnamefont{and}
  \bibinfo{author}{\bibfnamefont{O.}~\bibnamefont{Gunnarsson}},
  \bibinfo{journal}{Rev. Mod. Phys.} \textbf{\bibinfo{volume}{61}},
  \bibinfo{pages}{689} (\bibinfo{year}{1989}).

\bibitem[{\citenamefont{Aulbur et~al.}(2000)\citenamefont{Aulbur, Jonsson, and
  Wilkins}}]{Aulbur00}
\bibinfo{author}{\bibfnamefont{W.}~\bibnamefont{Aulbur}},
  \bibinfo{author}{\bibfnamefont{L.}~\bibnamefont{Jonsson}}, \bibnamefont{and}
  \bibinfo{author}{\bibfnamefont{J.~W.} \bibnamefont{Wilkins}}, in
  \emph{\bibinfo{booktitle}{Solid State Physics}}, edited by
  \bibinfo{editor}{\bibfnamefont{H.}~\bibnamefont{Ehrenreich}}
  \bibnamefont{and} \bibinfo{editor}{\bibfnamefont{F.}~\bibnamefont{Spaepen}}
  (\bibinfo{publisher}{Academic Press}, \bibinfo{address}{San Diego San
  Francisco New York Boston London Sydney Tokyo}, \bibinfo{year}{2000}),
  vol.~\bibinfo{volume}{54}, p.~\bibinfo{pages}{1}.

\bibitem[{\citenamefont{Sandu and Kirk}(2005)}]{Sandu05}
\bibinfo{author}{\bibfnamefont{T.}~\bibnamefont{Sandu}} \bibnamefont{and}
  \bibinfo{author}{\bibfnamefont{W.~P.} \bibnamefont{Kirk}},
  \bibinfo{journal}{Phys. Rev. B} \textbf{\bibinfo{volume}{72}},
  \bibinfo{pages}{073204} (\bibinfo{year}{2005}).

\bibitem[{\citenamefont{Harrison}(1980)}]{8}
\bibinfo{author}{\bibfnamefont{W.~A.} \bibnamefont{Harrison}},
  \emph{\bibinfo{title}{Electronic Structure and the Properties of Solids}}
  (\bibinfo{publisher}{Freeman}, \bibinfo{address}{San Francisco},
  \bibinfo{year}{1980}).

\bibitem[{\citenamefont{Chadi}(1977)}]{7}
\bibinfo{author}{\bibfnamefont{D.~J.} \bibnamefont{Chadi}},
  \bibinfo{journal}{Phys. Rev. B} \textbf{\bibinfo{volume}{16}},
  \bibinfo{pages}{790} (\bibinfo{year}{1977}).

\bibitem[{\citenamefont{Fleszar and Hanke}(2000)}]{9}
\bibinfo{author}{\bibfnamefont{A.}~\bibnamefont{Fleszar}} \bibnamefont{and}
  \bibinfo{author}{\bibfnamefont{W.}~\bibnamefont{Hanke}},
  \bibinfo{journal}{Phys. Rev. B} \textbf{\bibinfo{volume}{62}},
  \bibinfo{pages}{2466} (\bibinfo{year}{2000}).

\bibitem[{\citenamefont{Hybertsen}(1994)}]{11}
\bibinfo{author}{\bibfnamefont{M.~S.} \bibnamefont{Hybertsen}},
  \bibinfo{journal}{Phys. Rev. Lett.} \textbf{\bibinfo{volume}{72}},
  \bibinfo{pages}{1514} (\bibinfo{year}{1994}).

\bibitem[{\citenamefont{Gell et~al.}(1986)\citenamefont{Gell, Nino, Jaros, and
  Herbert}}]{12}
\bibinfo{author}{\bibfnamefont{M.}~\bibnamefont{Gell}},
  \bibinfo{author}{\bibfnamefont{D.}~\bibnamefont{Nino}},
  \bibinfo{author}{\bibfnamefont{M.}~\bibnamefont{Jaros}}, \bibnamefont{and}
  \bibinfo{author}{\bibfnamefont{D.~C.} \bibnamefont{Herbert}},
  \bibinfo{journal}{Phys. Rev. B} \textbf{\bibinfo{volume}{34}},
  \bibinfo{pages}{2416} (\bibinfo{year}{1986}).

\bibitem[{\citenamefont{Schulman and Chang}(1985)}]{13}
\bibinfo{author}{\bibfnamefont{J.~N.} \bibnamefont{Schulman}} \bibnamefont{and}
  \bibinfo{author}{\bibfnamefont{Y.~C.} \bibnamefont{Chang}},
  \bibinfo{journal}{Phys. Rev. B} \textbf{\bibinfo{volume}{31}},
  \bibinfo{pages}{2056} (\bibinfo{year}{1985}).

\bibitem[{\citenamefont{Klimeck et~al.}(2000)\citenamefont{Klimeck, Bowen,
  Boykin, Salazar-Lazaro, Cwik, and Stoica}}]{Klimeck00}
\bibinfo{author}{\bibfnamefont{G.}~\bibnamefont{Klimeck}},
  \bibinfo{author}{\bibfnamefont{R.~C.} \bibnamefont{Bowen}},
  \bibinfo{author}{\bibfnamefont{T.~B.} \bibnamefont{Boykin}},
  \bibinfo{author}{\bibfnamefont{C.}~\bibnamefont{Salazar-Lazaro}},
  \bibinfo{author}{\bibfnamefont{T.~A.} \bibnamefont{Cwik}}, \bibnamefont{and}
  \bibinfo{author}{\bibfnamefont{A.}~\bibnamefont{Stoica}},
  \bibinfo{journal}{Supperlatt. Microstruct.} \textbf{\bibinfo{volume}{27}},
  \bibinfo{pages}{77} (\bibinfo{year}{2000}).

\bibitem[{\citenamefont{Harrison et~al.}(1978)\citenamefont{Harrison, Kraut,
  Waldrop, and Grant}}]{Harrison78}
\bibinfo{author}{\bibfnamefont{W.~A.} \bibnamefont{Harrison}},
  \bibinfo{author}{\bibfnamefont{E.~A.} \bibnamefont{Kraut}},
  \bibinfo{author}{\bibfnamefont{J.~R.} \bibnamefont{Waldrop}},
  \bibnamefont{and} \bibinfo{author}{\bibfnamefont{R.~W.} \bibnamefont{Grant}},
  \bibinfo{journal}{Phys. Rev. B} \textbf{\bibinfo{volume}{18}},
  \bibinfo{pages}{4402} (\bibinfo{year}{1978}).

\bibitem[{\citenamefont{Kroemer}(1987)}]{Kroemer87}
\bibinfo{author}{\bibfnamefont{H.}~\bibnamefont{Kroemer}}, \bibinfo{journal}{J.
  Cryst. Growth} \textbf{\bibinfo{volume}{81}}, \bibinfo{pages}{193}
  (\bibinfo{year}{1987}).

\bibitem[{\citenamefont{Eppenga}(1989)}]{Eppenga89}
\bibinfo{author}{\bibfnamefont{R.}~\bibnamefont{Eppenga}}, \bibinfo{journal}{Phys. Rev. B}
  \textbf{\bibinfo{volume}{40}}, \bibinfo{pages}{10402} (\bibinfo{year}{1989}).

\bibitem[{\citenamefont{Sala et~al.}(1999)\citenamefont{Della Sala, Di Carlo, Lugli,
  Bernardini, Fiorentini, Scholz, and Jancu}}]{Sala99}
\bibinfo{author}{\bibfnamefont{F.} \bibnamefont{Della Sala}},
  \bibinfo{author}{\bibfnamefont{A.} \bibnamefont{Di Carlo}},
  \bibinfo{author}{\bibfnamefont{P.}~\bibnamefont{Lugli}},
  \bibinfo{author}{\bibfnamefont{F.}~\bibnamefont{Bernardini}},
  \bibinfo{author}{\bibfnamefont{V.}~\bibnamefont{Fiorentini}},
  \bibinfo{author}{\bibfnamefont{R.}~\bibnamefont{Scholz}}, \bibnamefont{and}
  \bibinfo{author}{\bibfnamefont{J.~M.} \bibnamefont{Jancu}},
  \bibinfo{journal}{Appl. Phys. Lett.} \textbf{\bibinfo{volume}{74}},
  \bibinfo{pages}{2002} (\bibinfo{year}{1999}).

\bibitem[{\citenamefont{Bonfiglio et~al.}(2000)\citenamefont{Bonfiglio,
  Lomascolo, Traetta, Cingolani, Di Carlo, Della Sala, Lugli, Botchkarev, and
  Morkoc}}]{Bonfiglio00}
\bibinfo{author}{\bibfnamefont{A.}~\bibnamefont{Bonfiglio}},
  \bibinfo{author}{\bibfnamefont{M.}~\bibnamefont{Lomascolo}},
  \bibinfo{author}{\bibfnamefont{G.}~\bibnamefont{Traetta}},
  \bibinfo{author}{\bibfnamefont{R.}~\bibnamefont{Cingolani}},
  \bibinfo{author}{\bibfnamefont{A.} \bibnamefont{Di Carlo}},
  \bibinfo{author}{\bibfnamefont{F.} \bibnamefont{Della Sala}},
  \bibinfo{author}{\bibfnamefont{P.}~\bibnamefont{Lugli}},
  \bibinfo{author}{\bibfnamefont{A.}~\bibnamefont{Botchkarev}},
  \bibnamefont{and} \bibinfo{author}{\bibfnamefont{H.}~\bibnamefont{Morkoc}},
  \bibinfo{journal}{J. Appl. Phys.} \textbf{\bibinfo{volume}{87}},
  \bibinfo{pages}{2289} (\bibinfo{year}{2000}).

\bibitem[{\citenamefont{Saito and Ikoma}(1992)}]{14}
\bibinfo{author}{\bibfnamefont{T.}~\bibnamefont{Saito}} \bibnamefont{and}
  \bibinfo{author}{\bibfnamefont{T.}~\bibnamefont{Ikoma}},
  \bibinfo{journal}{Phys. Rev. B} \textbf{\bibinfo{volume}{45}},
  \bibinfo{pages}{1762} (\bibinfo{year}{1992}).

\bibitem[{\citenamefont{Laref et~al.}(2003)\citenamefont{Laref, Aourag,
  Belgoumene, and Tadjer}}]{Laref03}
\bibinfo{author}{\bibfnamefont{A.}~\bibnamefont{Laref}},
  \bibinfo{author}{\bibfnamefont{H.}~\bibnamefont{Aourag}},
  \bibinfo{author}{\bibfnamefont{B.}~\bibnamefont{Belgoumene}},
  \bibnamefont{and} \bibinfo{author}{\bibfnamefont{A.}~\bibnamefont{Tadjer}},
  \bibinfo{journal}{J. Appl. Phys.} \textbf{\bibinfo{volume}{94}},
  \bibinfo{pages}{5027} (\bibinfo{year}{2003}).

\bibitem[{\citenamefont{Laref et~al.}(2006)\citenamefont{Laref, Laref,
  Belgoumene, Bouhafs, Tadjer, and Aourag}}]{Laref06}
\bibinfo{author}{\bibfnamefont{A.}~\bibnamefont{Laref}},
  \bibinfo{author}{\bibfnamefont{S.}~\bibnamefont{Laref}},
  \bibinfo{author}{\bibfnamefont{B.}~\bibnamefont{Belgoumene}},
  \bibinfo{author}{\bibfnamefont{B.}~\bibnamefont{Bouhafs}},
  \bibinfo{author}{\bibfnamefont{A.}~\bibnamefont{Tadjer}}, \bibnamefont{and}
  \bibinfo{author}{\bibfnamefont{H.}~\bibnamefont{Aourag}},
  \bibinfo{journal}{J. Appl. Phys.} \textbf{\bibinfo{volume}{99}},
  \bibinfo{pages}{043702} (\bibinfo{year}{2006}).

\bibitem[{\citenamefont{Wang and Ting}(1995)}]{15}
\bibinfo{author}{\bibfnamefont{E.~G.} \bibnamefont{Wang}} \bibnamefont{and}
  \bibinfo{author}{\bibfnamefont{C.~S.} \bibnamefont{Ting}},
  \bibinfo{journal}{Phys. Rev. B} \textbf{\bibinfo{volume}{51}},
  \bibinfo{pages}{9791} (\bibinfo{year}{1995}).

\bibitem[{\citenamefont{Wang}(1996)}]{16}
\bibinfo{author}{\bibfnamefont{E.~G.} \bibnamefont{Wang}},
  \bibinfo{journal}{Appl. Surf. Sci.} \textbf{\bibinfo{volume}{105/105}},
  \bibinfo{pages}{626} (\bibinfo{year}{1996}).

\bibitem[{\citenamefont{Sze}(1980)}]{17}
\bibinfo{author}{\bibfnamefont{S.~M.} \bibnamefont{Sze}},
  \emph{\bibinfo{title}{Physics of Semiconductor Devices}}
  (\bibinfo{publisher}{John Wiley and Sons}, \bibinfo{address}{New
  York-Chichester-Brisbane-Toronto}, \bibinfo{year}{1980}),
  \bibinfo{edition}{2nd} ed.

\bibitem[{\citenamefont{Boykin}(1995)}]{Boykin95}
\bibinfo{author}{\bibfnamefont{T.~B.} \bibnamefont{Boykin}},
  \bibinfo{journal}{Phys. Rev. B} \textbf{\bibinfo{volume}{52}},
  \bibinfo{pages}{16317} (\bibinfo{year}{1995}).

\bibitem[{\citenamefont{Graf and Vogl}(1995)}]{18}
\bibinfo{author}{\bibfnamefont{M.}~\bibnamefont{Graf}} \bibnamefont{and}
  \bibinfo{author}{\bibfnamefont{P.}~\bibnamefont{Vogl}},
  \bibinfo{journal}{Phys. Rev. B} \textbf{\bibinfo{volume}{51}},
  \bibinfo{pages}{4940} (\bibinfo{year}{1995}).

\bibitem[{\citenamefont{Boykin and Vogl}(2002)}]{Boykin02}
\bibinfo{author}{\bibfnamefont{T.~B.} \bibnamefont{Boykin}} \bibnamefont{and}
  \bibinfo{author}{\bibfnamefont{P.}~\bibnamefont{Vogl}},
  \bibinfo{journal}{Phys. Rev. B} \textbf{\bibinfo{volume}{65}},
  \bibinfo{pages}{035202} (\bibinfo{year}{2002}).

\bibitem[{\citenamefont{Boykin et~al.}(2001)\citenamefont{Boykin, Bowen, and
  Klimeck}}]{Boykin02b}
\bibinfo{author}{\bibfnamefont{T.~B.} \bibnamefont{Boykin}},
  \bibinfo{author}{\bibfnamefont{R.~C.} \bibnamefont{Bowen}}, \bibnamefont{and}
  \bibinfo{author}{\bibfnamefont{G.}~\bibnamefont{Klimeck}},
  \bibinfo{journal}{Phys. Rev. B} \textbf{\bibinfo{volume}{63}},
  \bibinfo{pages}{245314} (\bibinfo{year}{2001}).

\bibitem[{\citenamefont{Pedersen et~al.}(2001)\citenamefont{Pedersen, Pedersen,
  and Kriestensen}}]{Pedersen01}
\bibinfo{author}{\bibfnamefont{T.~G.} \bibnamefont{Pedersen}},
  \bibinfo{author}{\bibfnamefont{K.}~\bibnamefont{Pedersen}}, \bibnamefont{and}
  \bibinfo{author}{\bibfnamefont{T.~B.} \bibnamefont{Kriestensen}},
  \bibinfo{journal}{Phys. Rev. B} \textbf{\bibinfo{volume}{63}},
  \bibinfo{pages}{201101(R)} (\bibinfo{year}{2001}).

\bibitem[{\citenamefont{Cruz et~al.}(1999)\citenamefont{Cruz, Beltran, Wang,
  Taguena-Martinez, and Rubo}}]{21}
\bibinfo{author}{\bibfnamefont{M.}~\bibnamefont{Cruz}},
  \bibinfo{author}{\bibfnamefont{M.~R.} \bibnamefont{Beltran}},
  \bibinfo{author}{\bibfnamefont{C.}~\bibnamefont{Wang}},
  \bibinfo{author}{\bibfnamefont{J.}~\bibnamefont{Taguena-Martinez}},
  \bibnamefont{and} \bibinfo{author}{\bibfnamefont{Y.~G.} \bibnamefont{Rubo}},
  \bibinfo{journal}{Phys. Rev. B} \textbf{\bibinfo{volume}{59}},
  \bibinfo{pages}{15381} (\bibinfo{year}{1999}).

\bibitem[{\citenamefont{Sandu}(2005)}]{Sandu2005}
\bibinfo{author}{\bibfnamefont{T.}~\bibnamefont{Sandu}},
  \bibinfo{journal}{Phys. Rev. B} \textbf{\bibinfo{volume}{72}},
  \bibinfo{pages}{125105} (\bibinfo{year}{2005}).

\bibitem[{\citenamefont{Chadi and Cohen}(1973)}]{19}
\bibinfo{author}{\bibfnamefont{D.~J.} \bibnamefont{Chadi}} \bibnamefont{and}
  \bibinfo{author}{\bibfnamefont{M.~L.} \bibnamefont{Cohen}},
  \bibinfo{journal}{Phys. Rev. B} \textbf{\bibinfo{volume}{8}},
  \bibinfo{pages}{5747} (\bibinfo{year}{1973}).

\bibitem[{\citenamefont{Ren and Dow}(1988)}]{20}
\bibinfo{author}{\bibfnamefont{S.~Y.} \bibnamefont{Ren}} \bibnamefont{and}
  \bibinfo{author}{\bibfnamefont{J.~D.} \bibnamefont{Dow}},
  \bibinfo{journal}{Phys. Rev. B} \textbf{\bibinfo{volume}{38}},
  \bibinfo{pages}{1999} (\bibinfo{year}{1988}).

\end{thebibliography}

\end{document}